\documentclass[aps,twocolumn,showpacs,prl]{revtex4}
\usepackage[dvips]{graphicx}
\usepackage[english]{babel}
\usepackage[T1]{fontenc}
\usepackage{mathrsfs}
\usepackage[tbtags]{amsmath}
\usepackage{amssymb}
\usepackage{amsxtra}
\usepackage{amsopn}
\usepackage{latexsym}
\usepackage[mathcal]{eucal}
\usepackage{mathtools}

\newcommand{\BE}{\begin{equation}}
\newcommand{\EE}{\end{equation}}
\newcommand{\BA}{\begin{align}}
\newcommand{\EA}{\end{align}}

\begin{document}

\title{Universal scaling of gluon and ghost propagators in the infrared}

\author{Fabio Siringo}

\affiliation{Dipartimento di Fisica e Astronomia 
dell'Universit\`a di Catania,\\ 
INFN Sezione di Catania,
Via S.Sofia 64, I-95123 Catania, Italy}

\date{\today}

\begin{abstract}
A universal behavior is predicted for ghost and gluon propagators in the infrared.
The universal behavior is shown to be a signature of a
one-loop approximation and emerges naturally by the massive expansion that
predicts universal analytical functions for the inverse dressing functions
that do not depend on any parameter or color number. 
By a scaling of units and by adding an integration constant,
all lattice data, for different color numbers (and even quark content for the ghosts), 
collapse on the same universal curves predicted by the massive expansion.
\end{abstract}

%Keywords:
%QCD, Propagators, Scaling, Massive expansion


%\\ Keywords:
%QCD, Propagators,  Dispersion relations, Confinement.}

%12.38.Bx       Perturbative calculations
%12.38.Lg	Other nonperturbative calculations (QCD)
%12.38.Aw	General properties of QCD (dynamics, confinement, etc.)
%14.70.Dj	Gluons
%11.15.Tk       Other nonperturbative techniques (gauge field theories)

%12.38.Gc	Lattice QCD calculations (see also 11.15.Ha Lattice gauge theory)
%11.10.Ef       Field Theory: Lagrangian and Hamiltonian approaches
%11.15.-q	Gauge field theories
%12.20.-m       QED
%11.15.Bt       General properties of perturbation theory (gauge theory)
%12.38.-t	Quantum chromodynamics

\maketitle

In the last years, substantial progresses have been made in the knowledge of the elementary
two-point functions of QCD. In the Euclidean space and Landau gauge, 
the predictions of non-perturbative numerical 
methods\cite{aguilar10,alkofer,huber14,huber15g,huber15b,reinhardt14,gep2,varqed,varqcd,genself}
are in fair agreement with the results of lattice 
simulations\cite{bogolubsky,twoloop,dudal,binosi12,oliveira12,burgio15,bowman04,bowman05,su2glu,su2gho,duarte}, 
pointing towards a decoupled scenario with
a finite gluon propagator in the infrared. Moreover, deep in the infrared, by a massive expansion for the
exact Lagrangian\cite{ptqcd0,ptqcd}, explicit analytical expressions have been reported 
for the propagators\cite{ptqcd,ptqcd2,analyt}, from first principles.
At one-loop, the optimized expansion is in very good agreement with the lattice data in the Euclidean space\cite{ptqcd2} 
and provides a direct and simple way to explore the analytic properties of the propagators 
in Minkowski space\cite{analyt,spectral}.

Nevertheless,
even in the  case of pure $SU(N)$ Yang-Mills theory, the role played by the color number $N$ has not been
fully clarified yet. A comparison of the propagators was made by other authors before\cite{cucchieri07}
for different values of $N$. While some universal scaling was predicted by solution of a truncated set of Schwinger-Dyson
equations, a quantitatively different result was found on the lattice for $SU(2)$ and $SU(3)$, indicating a
similar behavior but only qualitative agreement for different values of $N$\cite{cucchieri07}.

In this brief note, we show that the agreement can be made {\it quantitative} by scaling the inverse dressing 
functions and adding an integration constant. 
In the infrared, all data are shown to collapse on the same curve by tuning the additive constant 
and by scaling the energy units, confirming a universal behaviour predicted by the massive 
expansion at one-loop\cite{ptqcd,ptqcd2,analyt}.
On the other hand, since higher loops would spoil the same prediction, the actual existence of the universal
scaling seems to be an indirect proof that the neglected higher order terms are very small in the optimized expansion.
Moreover, at one-loop the ghosts are decoupled from the quarks in the loop expansion, predicting the same
universal behavior even for unquenched data, irrespective of the number of quarks\cite{analyt}. 
Thus, the universal scaling of the
unquenched data would provide a further stringent test for the one-loop approximation of Ref.\cite{ptqcd2}.

The universal scaling emerges quite naturally by assuming that a generic one-loop approximation can be used deep in the
infrared, provided that the gluon mass arises from the loops and there is no mass at tree level. 

Of course, it does not need to be the standard one-loop expansion which breaks down below the low energy scale of QCD.
In principle, we can hypothesize the existence of a generic expansion in powers of an effective coupling, 
small enough to
give a reliable one-loop approximation even in the infrared.
All non-perturbative effects would be inside the definition
of that effective coupling of course and any link to the UV would require a full knowledge of the flow of 
the running coupling. A simple example is provided by the massive expansion of Ref.\cite{ptqcd2} 
that remains meaningful even in the limit $p\to 0$ where it gives a vanishing effective coupling 
in perfect agreement with the data of lattice simulations in the Landau gauge. 

Next, we must assume that the 
color number $N$ appears only as an argument of the effective coupling. 
That is what happens at one loop for the standard perturbation theory and for the 
massive expansion that share the same loop structure. Actually, neglecting RG effects, in a fixed-coupling scheme
the number $N$ is just a factor of the bare coupling at one loop. 
That is a trivial consequence of the existence of only one 
bare coupling in the Lagrangian, with no other free parameters. 
It would not be satisfied if the Lagrangian were changed by inclusion of spurious terms
or masses, as it happens for some massive models where the gluon mass is added by hand to the Lagrangian.
Moreover, deviations are expected in the UV where the RG running of the coupling cannot be neglected.
By a fixed coupling, the massive expansion provides a very good description of the data below 2 GeV, so that
the hypotheses are satisfied in the infrared. 

Denoting by $\alpha$ the effective coupling, assumed to be a function of $N$, the gluon and ghost self energies 
can be written in powers of $\alpha$ as
\BE
\Sigma( p)=\alpha \Sigma^{(1)}(p)+\alpha^2 \Sigma^{(2)}(p,N)+\cdots
\label{S}
\EE
where in general, we do not need to rule out an explicit dependence on $N$ for $\Sigma^{(2)}$ and higher-order terms.
While this is the structure of gluon and ghost self energies in the massive expansion of Ref.\cite{ptqcd2}, 
the following argument holds for any theory that has the same one-loop structure.

Any acceptable theory must also predict a finite gluon propagator in the IR, giving a mass scale
$m^2=\Delta (0)^{-1}$ which is arbitrary because of the arbitrary renormalization of the propagator. 
Since the Lagrangian is scaleless, the mass $m$ can only be determined by the phenomenology. 
For instance, in the massive expansion the mass scale is provided by an arbitrary gluon mass in the loops. 

Then, on general grounds, the self energy
can be written as
\BE
\frac{\Sigma( p)}{\alpha p^2}= -F(p^2/m^2) +{\cal O} (\alpha)
\label{S2}
\EE
where the adimensional function $F(s)$ is given by the one-loop self energy
\BE
F(p^2/m^2)=-\frac{\Sigma^{(1)}}{p^2}
\label{Fdef}
\EE
and can only depend on the ratio $s=p^2/m^2$.
The same argument applies to gluons and ghosts so that we can denote by $\Sigma$ the generic self energy
and by $\Delta$ the generic propagator. 
The exact gluon or ghost propagator $\Delta(p)$ can be written as
\BE
\Delta(p)=\frac{Z}{p^2-\Sigma(p)}=\frac{J(p)}{p^2}
\label{G}
\EE
where $Z$ is an arbitrary renormalization constant and $J(p)$ is the exact dressing function. 
As usual, at one loop, we can write $Z$ as the product of a finite renormalization constant $z$ times
a diverging factor $1+\alpha\>\delta Z$, so that the dressing function reads
\BE
z\>J(p)^{-1}=1+\alpha\left[F(p^2/m^2)-\delta Z\right]+{\cal O}(\alpha^2).
\label{chi}
\EE
Here, the divergent part of $\delta Z$ cancels the divergence of the one-loop self energy, yielding
a finite result. From now on let us ignore the diverging terms (that cancel) and assume that the
function $F(s)$ contains only the finite part, that is defined up to an additive (finite) 
renormalization constant $\delta Z$.
Eq.(\ref{chi}) already shows that the inverse dressing function $J^{-1}$ is entirely determined by the
adimensional function $F(s)$ up to a factor and an additive renormalization constant.
Moreover, we can divide by $\alpha$ and absorb the coupling in the arbitrary factor $z$ yielding
\BE
z\>J(p)^{-1}=F(p^2/m^2)+F_0+{\cal O}(\alpha).
\label{chi2}
\EE
where the new constant $F_0$ is the sum of all the constant terms.
If the higher order terms can be neglected, the dressing
functions are determined by the adimensional universal function $F(s)$ that does not depend on any parameter.
The remarkable result of Eq.(\ref{chi2}) requires that no tree-level mass is present in the self energy in
Eq.(\ref{G}), otherwise a term $1/(\alpha s)$ would be added to $F(s)$ and a dependence on $N$ would appear
in the dressing functions. In other words, the leading mass term must come from loops and must be of
order $\alpha$. Again, that condition is satisfied in the massive expansion of Refs.\cite{ptqcd,ptqcd2} but is not
met by other massive models.

We can associate a second energy scale $\mu$ to the finite renormalization constant $\delta Z$. For instance,
we can demand that $J(\mu)=z$ and obtain from Eq.(\ref{chi}) that $\delta Z=F(\mu^2/m^2)$. In that sense,
the constant $\delta Z$ is associated to the arbitrary choice of the subtraction point $\mu$ and Eq.(\ref{chi})
would take the more familiar aspect
\BE
z\>J(p)^{-1}=1+\alpha\left[F(p^2/m^2)-F(\mu^2/m^2)\right]+{\cal O}(\alpha^2).
\label{chi3}
\EE
The existence of two energy scales gives a spurious parameter given by the ratio $\mu/ m$. Once the units are
fixed by the phenomenology, the outcome of the theory should not depend on that parameter, but it does,
because of the approximations. The expansion can be optimized by looking for the best ratio that minimizes the
effects of higher loops. That is equivalent to taking the best value of $F_0$ in Eq.(\ref{chi2}). Then, the 
choice of that constant can be seen as a variational 
choice of the best subtraction point\cite{stevenson81,stevenson13,stevenson16}.
When optimized
by that method, the massive expansion provides a very good description of the lattice data in the Euclidean
space\cite{ptqcd2,analyt}.

Going back to the general result of Eq.(\ref{chi2}),
in order to illustrate its predictive power, let us take the first derivative of the inverse
dressing function $J^{-1}$. The derivative is entirely determined by the derivative of the universal
function $F(s)$. Then, scaling the units by the factors $m$ and $z$, all the lattice data must collapse on
the same curve, given by the derivative of $F(s)$. That should be more evident in a log-log plot where the curves
could be translated one on top of the other. However, since it is not easy to take the derivative for a set of
data points, we can integrate and get back Eq.(\ref{chi2}) where the constant $F_0$ now appears as an
integration constant. For any data set, any color number and bare coupling (and any quark content for the
ghost sector), three constants $x,y,z$ must exist such that the lattice inverse dressing function can be
written as
\BE
z\>J(p/x)^{-1}+y=F(p^2/m^2)+F_0+{\cal O}(\alpha)
\label{Jxyz}
\EE
thus collapsing on the same curve if higher order terms are negligible.

\begin{table}[ht]
\centering % used for centering table
\begin{tabular}{c c c c c c c c} % centered columns (8 columns)
\hline\hline %inserts double horizontal lines
Data set & $N$ & $N_f$ & $x$ &  $y$ & $z$ & $y^\prime$ & $z^\prime$   
\\ [0.5ex] % inserts table
%heading
\hline % inserts single horizontal line

Bogolubsky et al.\cite{bogolubsky} & 3 & 0 & 1 & 0 & 3.33 & 0 & 1.57 \\ % inserting body of the table
Duarte et al.\cite{duarte}         & 3 & 0 & 1.1 & -0.146 & 2.65 & 0.097 & 1.08 \\ [1ex] % [1ex] adds vertical space
\hline %inserts single line
Cucchieri-Mendes\cite{cucchieri08,cucchieri08b}
& 2 & 0 & 0.858 & -0.254 & 1.69 & 0.196 & 1.09 \\ [1ex] % [1ex] adds vertical space
\hline %inserts single line
Ayala et al.\cite{binosi12} & 3 & 0 & 0.933 & - & - & 0.045 & 1.17 \\ % inserting body of the table
Ayala et al.\cite{binosi12} & 3 & 2 & 1.04 & - & - & 0.045 & 1.28 \\ % inserting body of the table
Ayala et al.\cite{binosi12} & 3 & 4 & 1.04 & - & - & 0.045 & 1.28 \\[1ex]  
\hline %inserts single line
\end{tabular}
\label{table} % is used to refer this table in the text
\caption{Scaling constants $x$, $y$, $z$ (gluon) and $y^\prime$, $z^\prime$ (ghost) as defined 
in Eqs.(\ref{Jxyz}),(\ref{chixyz}). The constant shifts $F_0=-1.05$, $G_0=0.24$ and the mass 
$m=0.73$ GeV are optimized by requiring that $x=1$ and $y=y^\prime=0$ for the lattice 
data of Bogolubsky et al.\cite{bogolubsky}.}
\end{table}

\begin{figure}[t] 
\centering
\includegraphics[width=0.35\textwidth,angle=-90]{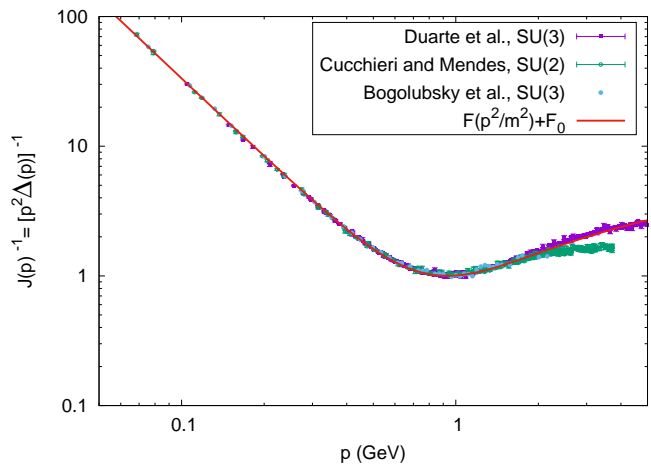}
\caption{The inverse gluon dressing function $z J^{-1}(p/x)+y$ scaled by the parameters $x$,$y$,$z$ of
Table I for each data set. 
The lattice data of Bogolubsky et al.\cite{bogolubsky} and Duarte et al.\cite{duarte}
for $SU(3)$ are compared with the data of Cucchieri and Mendes\cite{cucchieri08,cucchieri08b} for
$SU(2)$. The solid curve (red line) is the one-loop universal function $F(s)$ of Eq.(\ref{Jxyz}), 
evaluated by the massive expansion of Refs.\cite{ptqcd,ptqcd2} for $s=p^2/m^2$, $m=0.73$ GeV and shifted
by the constant $F_0=-1.05$.}
\label{fig1}
\end{figure}

\begin{figure}[t] 
\centering
\includegraphics[width=0.35\textwidth,angle=-90]{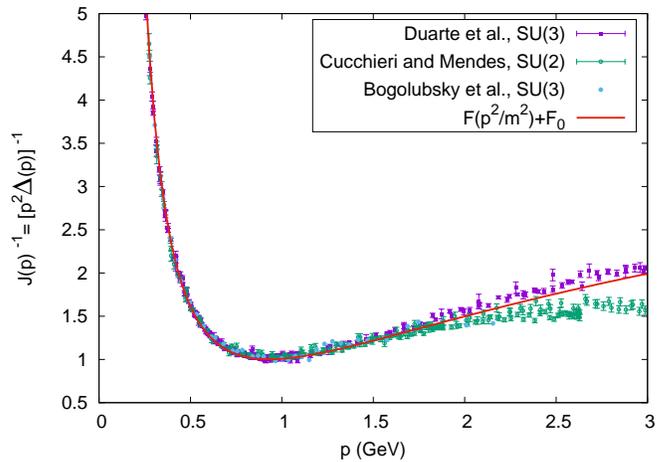}
\caption{Inverse gluon dressing function. The same content of Fig.\ref{fig1} is shown at a larger linear scale.}
\label{fig2}
\end{figure}

\begin{figure}[b] 
\centering
\includegraphics[width=0.35\textwidth,angle=-90]{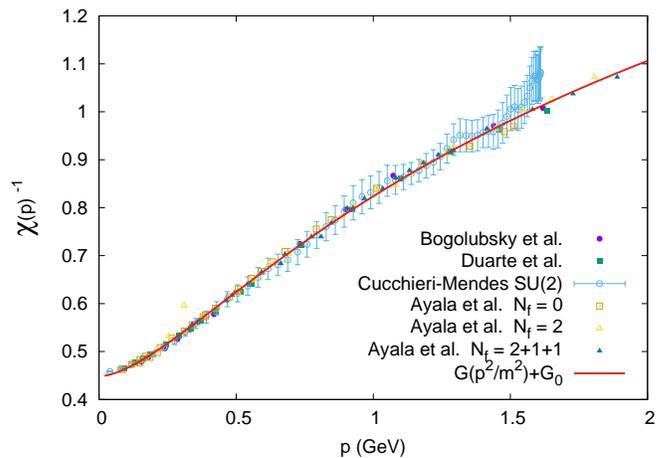}
\caption{The inverse ghost dressing function $z^\prime \chi^{-1}(p/x)+y^\prime$ scaled by the parameters 
$x$,$y^\prime$,$z^\prime$ of Table I for each data set. 
The lattice data of Bogolubsky et al.\cite{bogolubsky} and Duarte et al.\cite{duarte}
for $SU(3)$ are compared with the data of Cucchieri and Mendes\cite{cucchieri08,cucchieri08b} for
$SU(2)$ and with the unquenched data of  Ayala et al.\cite{binosi12} for full QCD with $N_f=2$ and
$N_f=2+1+1$.
The solid curve (red line) is the one-loop universal function $G(s)$ of Eq.(\ref{chixyz}), 
evaluated by the massive expansion of Refs.\cite{ptqcd,ptqcd2} for $s=p^2/m^2$, $m=0.73$ GeV and shifted
by the constant $G_0=0.24$.}
\label{fig3}
\end{figure}

\begin{figure}[t] 
\centering
\includegraphics[width=0.35\textwidth,angle=-90]{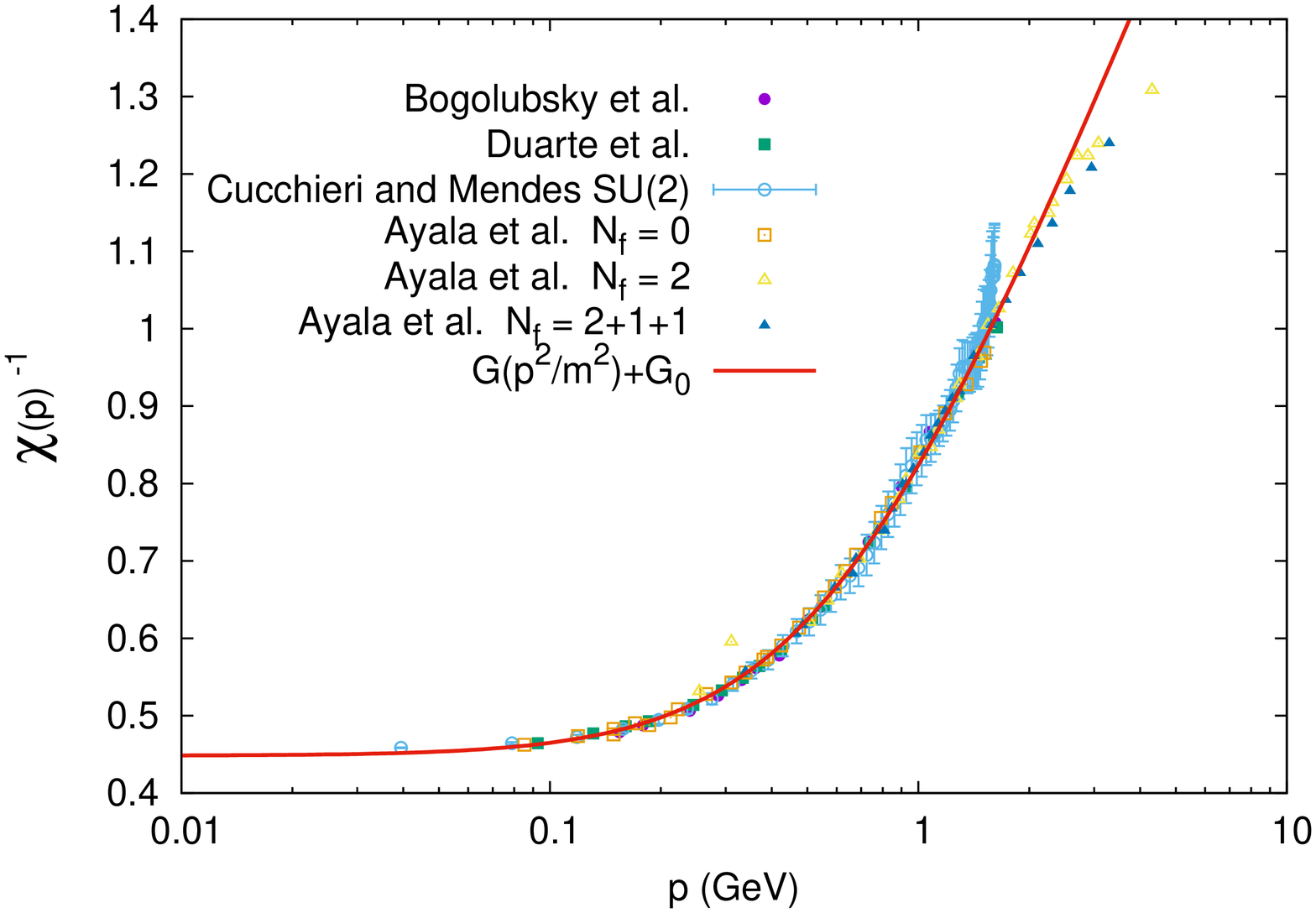}
\caption{Inverse ghost dressing function. The same content of Fig.\ref{fig3} is shown on
a wider range by a logarithmic scale.}
\label{fig4}
\end{figure}

Actually, as shown in Figs.\ref{fig1},\ref{fig2}, denoting now by $J(p)$ a set of lattice data points 
for the {\it gluon} dressing function, 
Eq.(\ref{Jxyz}) is very well satisfied in the infrared below 2 GeV with the renormalization constants $x,y,z$
of Table I. 
In Fig.\ref{fig1}
the lattice data points of Bogolubsky et al.\cite{bogolubsky} and Duarte et al.\cite{duarte}
for $SU(3)$ are compared with the data points of Cucchieri and Mendes\cite{cucchieri08,cucchieri08b} for
$SU(2)$. When scaled by the constants of Table I, the data collapse on a single curve that is very well
described by the one-loop universal function $F(s)$, evaluated by the massive expansion of Refs.\cite{ptqcd,ptqcd2}
and shown in the figure as a solid line for $m=0.73$ GeV and $F_0=-1.05$. The same data points are shown
in Fig.\ref{fig2} at a larger linear scale. Deviations occur above 2 GeV where RG effects must be considered
for a correct description of the UV limit\cite{ptqcd2}.

Denoting by $\chi(p)$ a set of lattice data points for the {\it ghost} dressing function, the
twin equation holds
\BE
z^\prime \>\chi(p/x)^{-1}+y^\prime=G(p^2/m^2)+G_0+{\cal O}(\alpha)
\label{chixyz}
\EE
where $G(s)$ is the ghost universal function that arises from Eq.(\ref{Fdef}) and $G_0$ is the corresponding integration
constant. We observe that the factor $x$ and the mass $m$, giving the energy units, must be the same for ghost and 
gluon within the same lattice simulation, as shown in Table I. Since no quark vertex is present in the
one-loop ghost self-energy, Eq.(\ref{chixyz}) must be satisfied
in the infrared even by the unquenched lattice data, as shown in Figs.\ref{fig3},\ref{fig4}.
In Fig.\ref{fig3}
the lattice data points of Bogolubsky et al.\cite{bogolubsky} and Duarte et al.\cite{duarte}
for $SU(3)$ are compared with the data points of Cucchieri and Mendes\cite{cucchieri08,cucchieri08b} for
$SU(2)$ and with the unquenched data of Ayala et al.\cite{binosi12} for full QCD with a fermion number $N_f=2$ and
$N_f=4$. Again, When scaled by the constants of Table I, the data collapse on a single curve that is very well
described by the one-loop universal function $G(s)$, evaluated by the massive expansion of Refs.\cite{ptqcd,ptqcd2}
and shown in the figure as a solid line for $m=0.73$ GeV and $G_0=0.24$. We observe that no change of units is
required going from the gluon to the ghost plot within the same data set, while minor differences in the
parameter $x$ are found between the data sets of different simulations, as a consequence of a slightly different
calibration of the physical units. Error bars are only shown for the $SU(2)$ set where they are quite larger than
the size of points. An overview of the same data is shown
in Fig.\ref{fig4} by a logarithmic scale. 

The universal scaling of the dressing functions is an indirect proof that higher order terms must add very small
contributions to the one-loop approximation. It would be interesting to extend the comparison to 
other one-loop massive models\cite{tissier10,tissier11} in order to explore the effects of spurious tree-level
mass terms.

The perfect agreement with the one-loop analytical
expressions of the massive expansion, gives us more confidence on the reliability of that method in the
infrared. We remark that the universal functions $F(s)$, $G(s)$ arise from first principles by the massive expansion
and their simple expressions\cite{ptqcd,ptqcd2} 
can be analytically continued to Minkowski space where they provide important physical
insights\cite{analyt}.  

The author is in debt to A. Cucchieri and O. Oliveira for sharing the data of their
lattice simulations.

\end{document}